\documentclass[pre,aps,superscriptaddress,showpacs,preprint]{revtex4-1}
\usepackage{amssymb,amsmath}
\usepackage{graphicx}
\usepackage{nicefrac}

\begin{document}

\title
{Micro-branching in mode-I fracture in a randomly perturbed lattice}
\author
{Shay I. Heizler}
\affiliation{Department of Physics, Bar-Ilan University,
 Ramat-Gan, IL52900 ISRAEL}
\affiliation{Department of Physics, Nuclear Research Center-Negev,
 P.O. Box 9001, Beer Sheva 84190, ISRAEL}
\author
{David A. Kessler}
\email{kessler@dave.ph.biu.ac.il}
\affiliation{Department of Physics, Bar-Ilan University,
 Ramat-Gan, IL52900 ISRAEL}
\author
{Yonatan S. Elbaz}
\affiliation{Department of Physics, Nuclear Research Center-Negev,
 P.O. Box 9001, Beer Sheva 84190, ISRAEL}
 

\begin{abstract}
We study mode-I fracture in lattices with noisy bonds. In contrast to previous attempts, by using a small parameter that
perturbs the force-law between the atoms in perfect lattices and using a 3-body force law, simulations reproduce the qualitative behavior of the beyond steady-state cracks in the high velocity regime,
including reasonable micro-branching. As far as the physical properties such as the structure factor $g(r)$, the radial or angular distributions,
these lattices share the physical properties of perfect lattices rather than that of an amorphous material
(e.g., the continuous random network model). A clear transition can be seen between steady-state cracks, where a single crack propagates in the midline of the sample and the regime of
unstable cracks, where micro-branches start to appear near the main crack, in line with previous experimental results. This is seen both in a honeycomb lattice
and a fully hexagonal lattice. This model reproduces the main physical features of propagating cracks in brittle materials, including the behavior of velocity as a function of
driving displacement and the increasing amplitude of oscillations of the electrical resistance. In addition, preliminary indications of  power-law behavior of the micro-branch shapes can be seen,
potentially reproducing one of the most intriguing experimental results of brittle fracture.

\end{abstract}

\maketitle
 
\section{Introduction and Brief Review}

An extensive experimental effort on mode-I (tensile) fracture in amorphous materials has been made in the last two
decades~\cite{ravi_chandar_knauss,ramulu_kobayashi,marder_jay1,marder_jay2,fineberg_sharon0,fineberg_sharon1,fineberg_sharon2,fineberg_sharon3,fineberg_sharon4,fineberg_sharon5,silicon,nature_ko}
(for a review, see~\cite{review}). In the high crack velocity regime, the simple picture of a rapid steady-state crack generated via a given driving displacement,
and exhibiting a given crack velocity (of order the Rayleigh surface wave speed) breaks down, and small micro-branches
start to appear next to the main crack~\cite{fineberg_sharon0,fineberg_sharon1,fineberg_sharon2,fineberg_sharon3,fineberg_sharon4,fineberg_sharon5,nature_ko}. Upon further Increase of
the loading, the microscopic branches transform to large macro-branches.

The experimental phenomena of the appearance of micro-branches has been the subject of an extensive theoretical effort as well. There have been several attempts within the framework of continuum
models, based on the
linear elasticity fracture mechanics (LEFM) theory~\cite{freund}.
Yoffe predicted that steady-state cracks will become unstable at a specific crack velocity $v_{\mathrm{cr}}\approxeq0.73c_R$, based on maximal stress considerations~\cite{yoffe}.
Another {\em ad hoc} attempt based on energy considerations predicted a critical velocity of $v_{\mathrm{cr}}\approxeq0.5c_R$~\cite{eshelby}.
However, the mode-I experiments have shown that the critical velocity is material-dependent, refuting those LEFM predictions. Several LEFM-based works predict a micro-branching
instability at a material-dependent critical velocity~\cite{adda_bedia}, however, the specific parameter that determines the specific critical
velocity, $\Gamma(v_{\mathrm{cr}})$ is an input parameter to the theory.

Additional efforts have been made to explore the micro-branching phenomena based on LEFM~\cite{bouchbinder_procaccia1,bouchbinder_procaccia2}. However, as these works  themselves argue,
although they recover some features of the micro-branching instability, some main predictions such as that the energy flow to the micro-branches drops immediately, are unphysical (the
micro-branch arrests immediately). These results raise the possibility that the micro-branches phenomena is a 3D-phenomena, while in 2D the high-velocity instability occurs only for extreme crack
velocities ($v_{\mathrm{cr}}>0.8c_R$), as the crack oscillates. This argument is based on experiments, both in biaxial mode-I crack in rubber~\cite{biaxial},
and pure mode-I experiments in gels~\cite{gels}, and  associated theoretical works~\cite{bouchbinder_procaccia3,bouchbinder1}. However 
 the mode-I experiments in PMMA seem to indicate that beyond a certain crack velocity, the crack structure that emerges is essentially two-dimensional~\cite{fineberg_sharon2}.
There has been some success in exploring the instability using phenomenological mesoscale approaches based on the phase-field~\cite{phase} or on
cohesive zones~\cite{finite_element,cohesive,peridynamics}; these however are difficult to quantitatively relate to an underlying microscopic picture.

The failure of the continuum theory (LEFM) has given rise to an extensive theoretical effort using atomistic lattice
models~\cite{slepyan,slepyan2,marderliu,mardergross,kess_lev1,kess_lev2,kess1,kess_lev3,pechenik,shay1,shay2} and
lattice simulations~\cite{marderliu,mardergross,fineberg_mar,mdsim1,mdsim2,mdsim3,kess_lev3,shay1,shay2,nature_old,nature}, where the inherent divergence of the
elastic fields near the crack tip of the continuum theories is tamed, due to the finite atomistic lattice scale. These models yield steady state cracks without any additional parameters,
once the inter-atomic forces are specified.
Both lattice models and simulations have shown and reproduced the
sharp transition between steady states cracks, where only the bonds on the midline of the lattice are broken, and the post-instability behavior, where beyond some critical velocity,
other bonds start to fail. This critical velocity was found to have a strong dependence ($0.3c_R<v_{cr}\leqslant c_R$) on the parameters of the potential, i.e. it is {\em material dependent}.
Although the lattice models yield the desired existence of a critical velocity, regarding the post-instability point behavior, the success is less impressive.
Mode-III (out-of-plane shear mode) simulations have
shown nice qualitative patterns of micro-branches~\cite{marderliu,kess_lev3}, similar mode-I simulations (the mode for which most of the experiments actually have been performed),
have failed to reproduce the qualitative patterns of micro-branches emerging near the main crack~\cite{fineberg_mar,shay1}.

Thus, several attempts have been made to try to simulate cracks in {\em amorphous materials}, in which the bulk of experiments have been performed (e.g., glass, PMMA, Homalite-100 etc.).
The first attempts used the classic binary-alloy model (using two different kinds of particles)~\cite{procaccia2} simulating an amorphous material
failed~\cite{falk_langer,falk,tsviki}; The crack always arrested.
Recently~\cite{shay3}, a new approach were presented based on a continuous random network model (CRN) for simulating amorphous materials~\cite{zacharainsen,www}. In this model the sample
looks like a distorted lattice, while each atom shares the same number of nearest neighbor atoms.  The CRN model yields both steady-state cracks
and the main features of the micro-branching instability, including the increasing size of the micro-branches  and the increasing oscillations in the electrical
resistance of the sample~\cite{shay3} with increasing external loading. Recently~\cite{procaccia2}, Dauchot, et al. succeeded in generating propagating steady-state cracks using the binary-alloy model, by going to the extreme brittle
limit, where the force falls rapidly to zero for very small strains.
However, no information was reported regarding the high-velocity instability. Moreover, the $g(r)$ (RDF)  generated from the amorphous model presented in~\cite{procaccia2} are
much less similar to the $g(r)$'s of real amorphous materials than those the CRN's model generates.

In this work we focus on trying to reproduce the successful results of the CRN model also in lattices (where the previous attempts have failed, as explained above). To do so, we let
the force-law between the atoms  vary slightly  by changing randomly the lattice scale $a$ between the atoms. As a result, the equilibrium locations of the
atoms are slightly changed from their pure lattice locations, according to the modified force law. By breaking the pure symmetry of the perfect lattice, we hope to obtain a realistic micro-branching
phenomena in these simple structures. The semi-quantitative behaviors of the micro-branches, such as the length of the micro-branches, should be less noisy than the corresponding results for CRN's.

\section{The Model and Main Methodology}

In our model, each bond (between atoms $i$ and $j$) has a specific characteristic equilibrium distance at which the force is zero. This distance $a_{i,j}$ is taken to vary slightly from the constant distance $a_0$
by a factor of $\epsilon_{i,j}$ which is drawn from a uniform distribution:
\begin{equation}
a_{i,j}=(1+\epsilon_{i,j})a_0,\qquad i=1,2,\dots,n_{\mathrm{atoms}}, j\in nn(i)
\end{equation}
where $\epsilon_{ij}\in[-b,b]$ and $b$ is a {\em constant} for a given lattice and in this work ranges between $0\leqslant b\leqslant 0.1$, $a_0=4$ and $nn(i)$ refers to the nearest-neighbors of site $i$.

Between each two atoms there is a piece-wise linear radial force (2-body force law) of the form:
\begin{equation}
\vec{f}^R_{i,j}=k_R(\vert\vec{r}_{i,j}\vert-a_{i,j})\theta_H\left(\varepsilon-\vert\vec{r}_{i,j}\vert\right)\hat{r}_{j,i},
\label{force_Radial}
\end{equation}
where the Heaviside step function $theta_H$ guarantees  that the force  drops immediately to zero when the distance between two atoms $\vert\vec{r}_{i,j}\vert$ reaches  a certain value $\varepsilon>a_{i,j}$ (the break of a bond).
In this work we set $\varepsilon=a_0+1$, and the units are chosen so that the spring constant $k_R$ is unity.
Potentially, in addition there is a 3-body force law that depends on the cosine of each of the angles, defined of course by:
\begin{equation}
\cos\theta_{i,j,k}=\frac{\vec{r}_{i,j}\cdot\vec{r}_{i,k}}{\vert\vec{r}_{i,j}\vert\vert\vec{r}_{i,k}\vert}
\label{cos_teta}
\end{equation}
In a honeycomb lattice there are three angles associated with each atom and in the hexagonal lattice there are six of them (we note that in the hexagonal lattice this choice is a little
bit arbitrary since there are in general additional optional angles for each atom). There is a certain preferred angle $\theta_C$ for which
the 3-body force law vanishes (in the honeycomb lattice we set $\theta_C=\nicefrac{2\pi}{3}$ and in the hexagonal lattice we set $\theta_C=\nicefrac{\pi}{3}$).
The 3-body force law that acts on the central atom (atom $i$) of each angle may expressed as:
\begin{align}
\label{force_teta}
& \vec{f}^{\theta}_{i,(j,k)}=k_{\theta}(\cos\theta_{i,j,k}-\cos\theta_C)\frac{\partial\cos\theta_{i,j,k}}{\partial\vec{r}_i}
\theta_H\left(\varepsilon-\vert\vec{r}_{i,j}\vert\right)\hat{r}_i=\\
& k_{\theta}(\cos\theta_{i,j,k}-\cos\theta_C)\left[\frac{\vec{r}_{i,j}+\vec{r}_{i,k}}{\vert\vec{r}_{i,j}\vert\vert\vec{r}_{i,k}\vert}+
\frac{\vec{r}_{j,i}(\vec{r}_{i,j}\cdot\vec{r}_{i,k})}{\vert\vec{r}_{i,j}\vert^3\vert\vec{r}_{i,k}\vert}+\right. \nonumber \\
& \left.\frac{\vec{r}_{k,i}(\vec{r}_{i,j}\cdot\vec{r}_{i,k})}{\vert\vec{r}_{i,j}\vert\vert\vec{r}_{i,k}\vert^3} \right]\theta_H\left(\varepsilon-\vert\vec{r}_{i,j}\vert\right), \nonumber
\end{align}
while the force  that is applied on the other two atoms (atoms $j,k$) may expressed as:
\begin{align}
\label{force_teta2}
& \vec{f}^{\theta}_{j,(i,k)}=k_{\theta}(\cos\theta_{i,j,k}-\cos\theta_C)\frac{\partial\cos\theta_{i,j,k}}{\partial\vec{r}_j}
\theta_H\left(\varepsilon-\vert\vec{r}_{i,j}\vert\right)\hat{r}_j=\\
& k_{\theta}(\cos\theta_{i,j,k}-\cos\theta_C)\left[\frac{\vec{r}_{k,i}}{\vert\vec{r}_{i,j}\vert\vert\vec{r}_{i,k}\vert}+
\frac{\vec{r}_{i,j}(\vec{r}_{i,j}\cdot\vec{r}_{i,k})}{\vert\vec{r}_{i,j}\vert^3\vert\vec{r}_{i,k}\vert} \right]\theta_H\left(\varepsilon-\vert\vec{r}_{i,j}\vert\right) \nonumber
\end{align}
Of course, the forces satisfy the relation:
$\vec{f}^{\theta}_{i,(j,k)}=-(\vec{f}^{\theta}_{j,(i,k)}+\vec{f}^{\theta}_{k,(i,j)})$.

In addition, we used a Kelvin-type viscoelastic force,  proportional to the relative velocity between the two atoms of the bond $\vec{v}_{i,j}$:
\begin{equation}
\vec{g}^R_{i,j}=\eta(\vec{v}_{i,j}\cdot\hat{r}_{i,j})\theta_H\left(\varepsilon-\vert\vec{r}_{i,j}\vert\right)\hat{r}_{i,j},
\label{viscous}
\end{equation}
with $\eta$  the viscosity parameter. The viscous force vanishes after the bond is broken.
Thus, the equation of motion of each atom is:
\begin{equation}
m_i\vec{\ddot{a}}_i=\sum_{j\in3p\;nn}\left(\vec{f}^R_{i,j}+\vec{g}^R_{i,j}\right)+\sum_{j,k\in3p\;nn}\vec{f}^{\theta}_{i,(j,k)}+\sum_{j\in6p\;nn}\vec{f}^{\theta}_{j,(i,k)},
\label{motion_equations}
\end{equation}
where $p=1$ for the honeycomb lattice and $p=2$ for the  hexagonal lattice ($nn$=nearest neighbors).  The masses $m_i$ can also be set to unity without loss of generality.

\begin{figure}
\centering{
\includegraphics*[width=7cm]{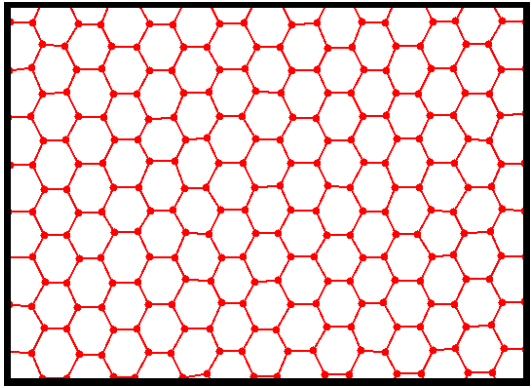}
}
\caption{A typical steady-state perturbed lattice grid using $b=10\%$.
}
\label{reshet}
\end{figure}

\begin{figure}
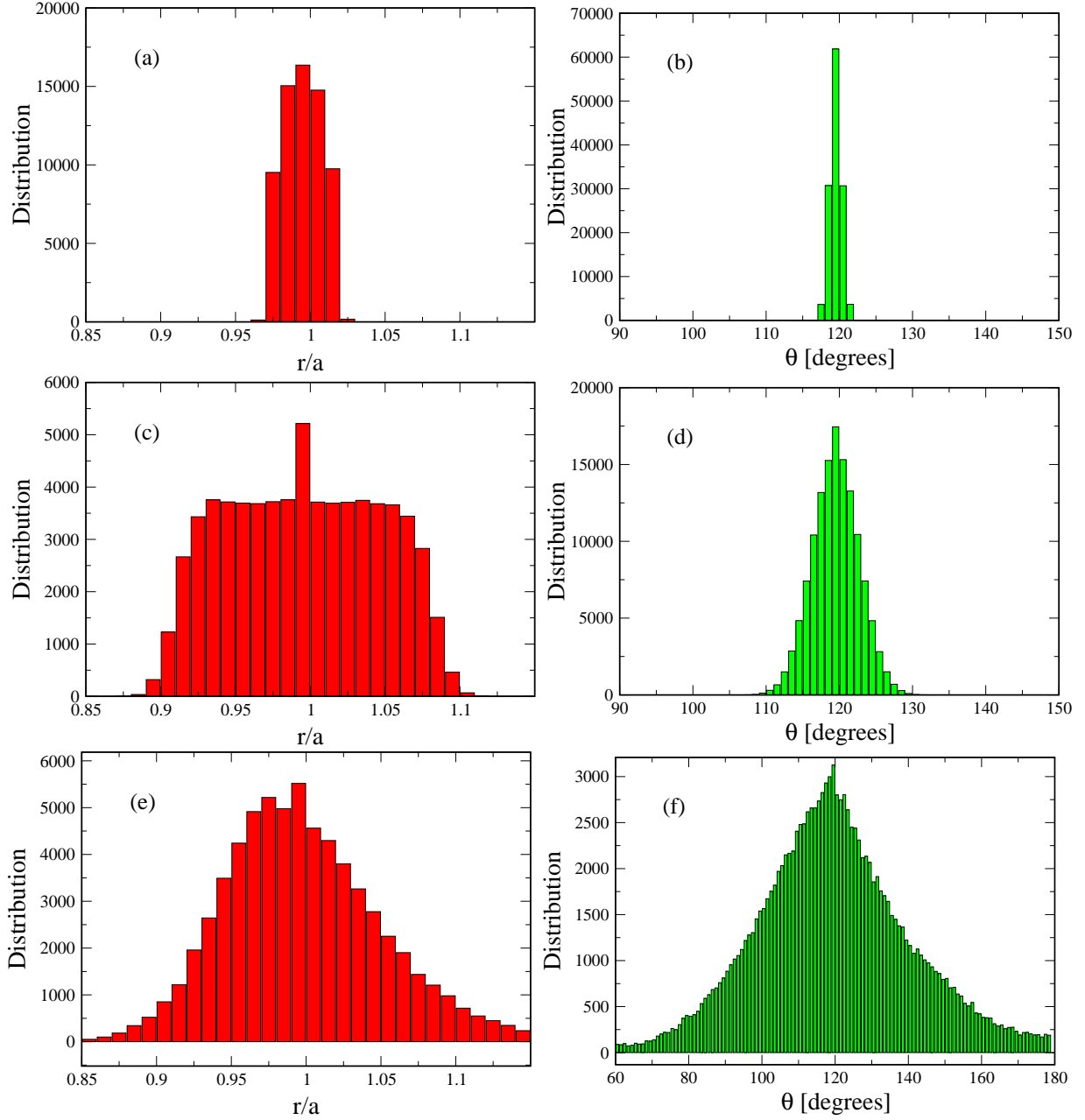

\centering{
\includegraphics*[width=8cm]{dist_2_5.eps}
\includegraphics*[width=8cm]{angle_2_5.eps}
\includegraphics*[width=8cm]{dist_10.eps}
\includegraphics*[width=8cm]{angle_10.eps}
\includegraphics*[width=8cm]{dist_crn.eps}
\includegraphics*[width=8cm]{angle_crn.eps}
}
\caption{The radial (a) and the angular (b) distributions for a perturbed lattice mesh with $b=2.5\%$. (c-d) Same for $b=10\%$. (e-f) Same for the CRN amorphous mesh.
}
\label{fig1}
\end{figure}

The main methodology is as follows: After choosing the random value $a_{i,j}$ for each bond, we allow the network to relax through a simple
molecular-dynamics Euler scheme, in accord with Eqs. (\ref{motion_equations}), with a non-zero $\eta$, until the total energy is minimized.
In Fig. \ref{reshet}, we can see an example of a perturbed honeycomb lattice with $b=10\%$, while in Fig. \ref{fig1} we can see the distribution of the radial distances
of the bonds for a perturbed lattice with $b=2.5\%$, $b=10\%$ along with the CRN, taken from~\cite{shay3}.

We can see the qualitative difference between the perturbed lattice mesh and the CRN. While the radial distributions for the perturbed lattice mesh are flat in the
range of $1\pm b$ and then drops immediately to zero (since the random distribution was taken to be flat in the range of $1\pm b$), the CRN has a long tails extending over
larger distances. In addition, the angular distributions are much narrower than the CRN's angular distribution.

\begin{figure}
\centering{
\includegraphics*[width=8cm]{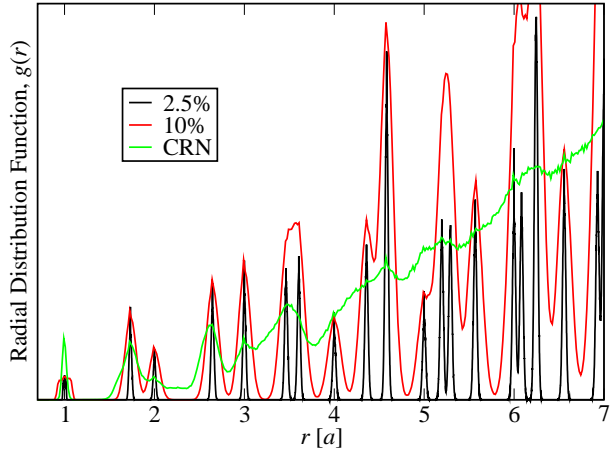}
}
\caption{The radial distribution function $g(r)$ as a function of $r$ (in lattice scale units) in the perturbed-lattice systems with different magnitude and in the 
continuous random network system. A major qualitative difference can be seen between the two networks.
}
\label{fig2}
\end{figure}

A powerful tool to check the character of the grid is of course the radial distribution function (RDF or $g(r)$). In Fig. \ref{fig2} we can see the RDFs of the perturbed
lattices and the CRN.
We can see again the qualitative difference between the meshes. While the CRN looks very much like a real amorphous material (see~\cite{shay3}), the RDFs of the
perturbed lattices look exactly like a pure lattice RDF (set of $\delta$-functions), only slightly  perturbed due to the random noise,
even for large $r$. Thus, the structure of the perturbed lattices is like a lattice material rather than an amorphous material.

After characterizing the initial lattices, we stretch the lattice under a mode-I tensile loading with a given constant strain using a given driving displacement $\Delta$,
and seed the system with an initial crack. We let the crack propagate via the same molecular dynamics Euler scheme that was introduced before. The lattice mesh we use contains
$162\cdot272\approx45,000$ ($N=80$ in the Slepyan model notation) atoms for the honeycomb lattice and $162\cdot408\approx65,000$ atoms for the hexagonal lattice.

\section{Honeycomb Lattice}

\begin{figure}
\centering{
\includegraphics*[width=8cm]{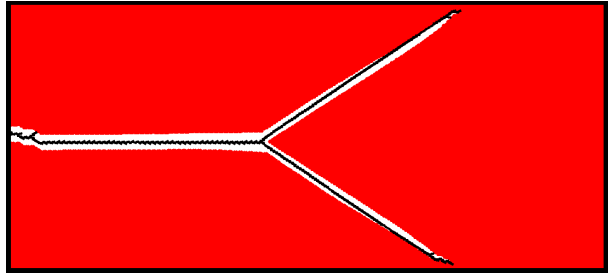}
}
\caption{The propagating crack snapshot using a pure honeycomb lattice using $\eta=0.25$ and $\nicefrac{\Delta}{\Delta_G}=2.7$. The crack bifurcates to two that travel to the end of the sample.
}
\label{pure_lattice_crack}
\end{figure}

Using $b=0$, i.e. an unperturbed honeycomb lattice, we obtain the well-known non-physical behavior of the crack above threshold. For small strains we get a perfect steady-state crack, while upon increasing the
driving displacement, the crack bifurcates to two macro-branches that propagate to the edges of the sample~\cite{shay1}. The same happens in the honeycomb lattice
including the 3-body force law, both with large-viscosity and with a negligible viscosity (Fig. \ref{pure_lattice_crack}).

\begin{figure}
\centering{
\includegraphics*[width=7cm]{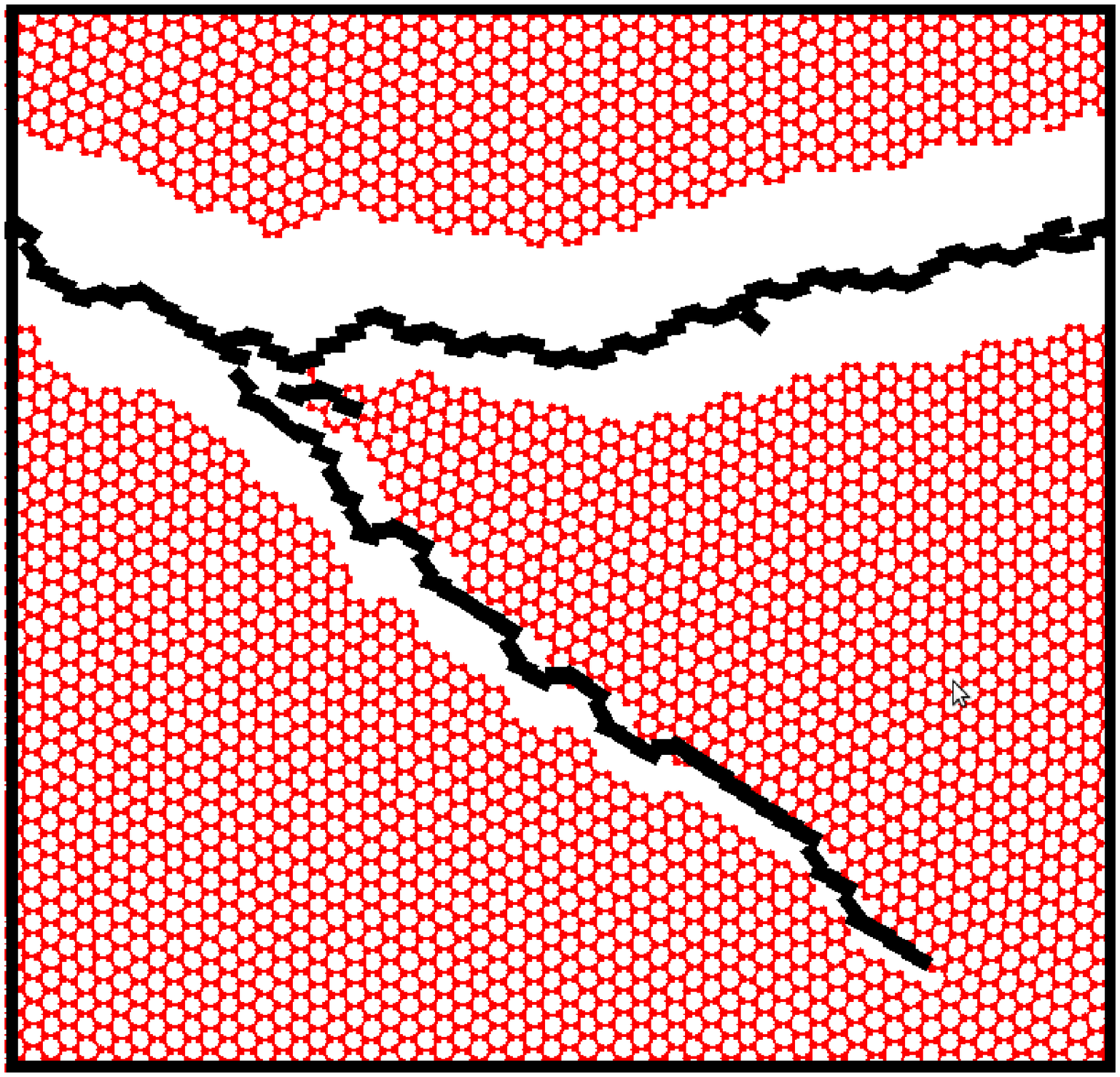}
\includegraphics*[width=7cm]{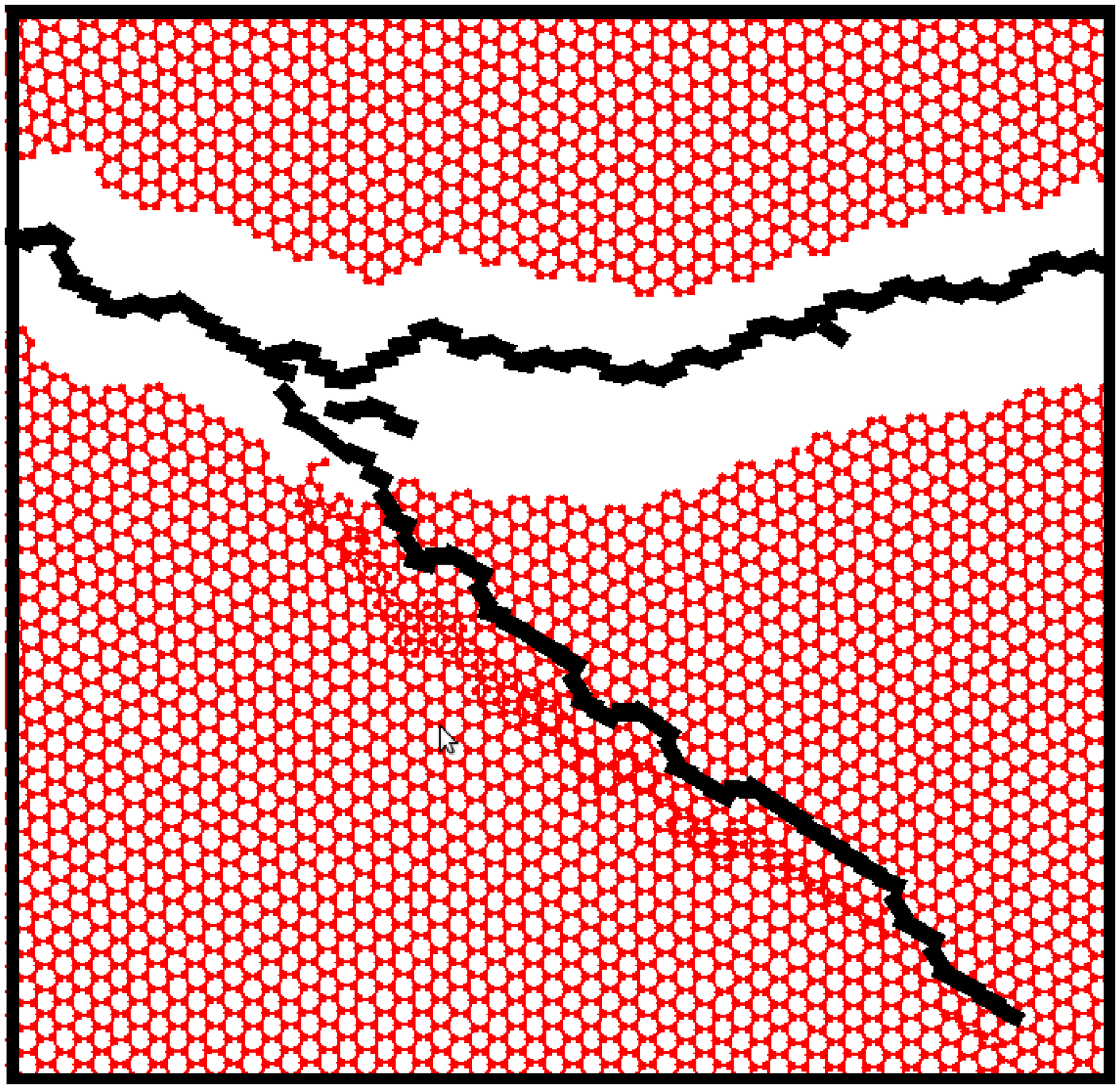}
}
\caption{(a) A propagating crack snapshot using a perturbed honeycomb lattice model using $b=10\%$, $\eta=2$ and $\nicefrac{\Delta}{\Delta_G}=3.8$. (b) A snapshot of the overlapping problem.
After cracking, the two pieces of the cracked lattice overlap, because fracture is an irreversible process in this model. 
}
\label{beehive_crack1}
\end{figure}

\begin{figure}
\centering{
\includegraphics*[width=7cm]{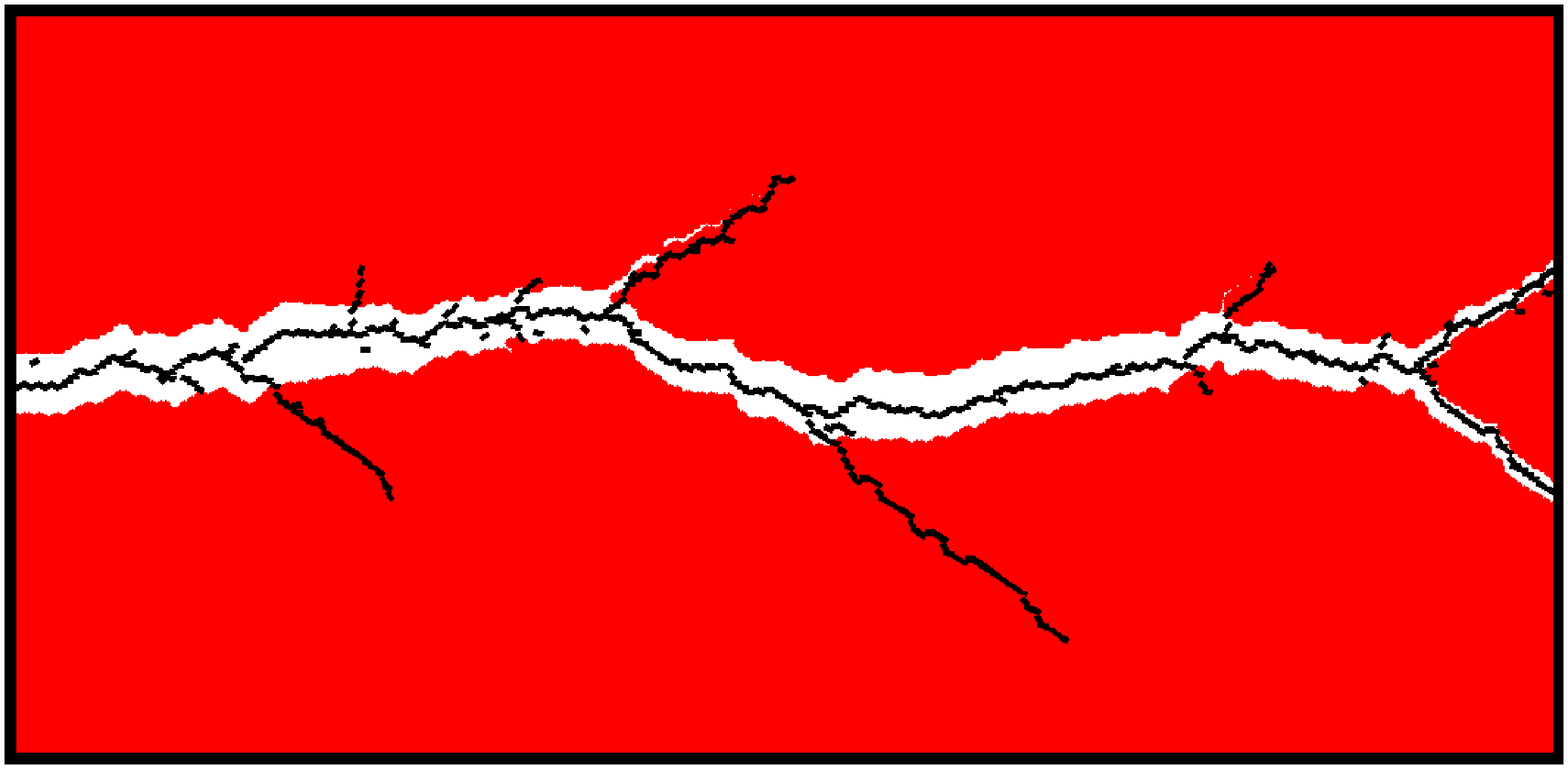}
\includegraphics*[width=7cm]{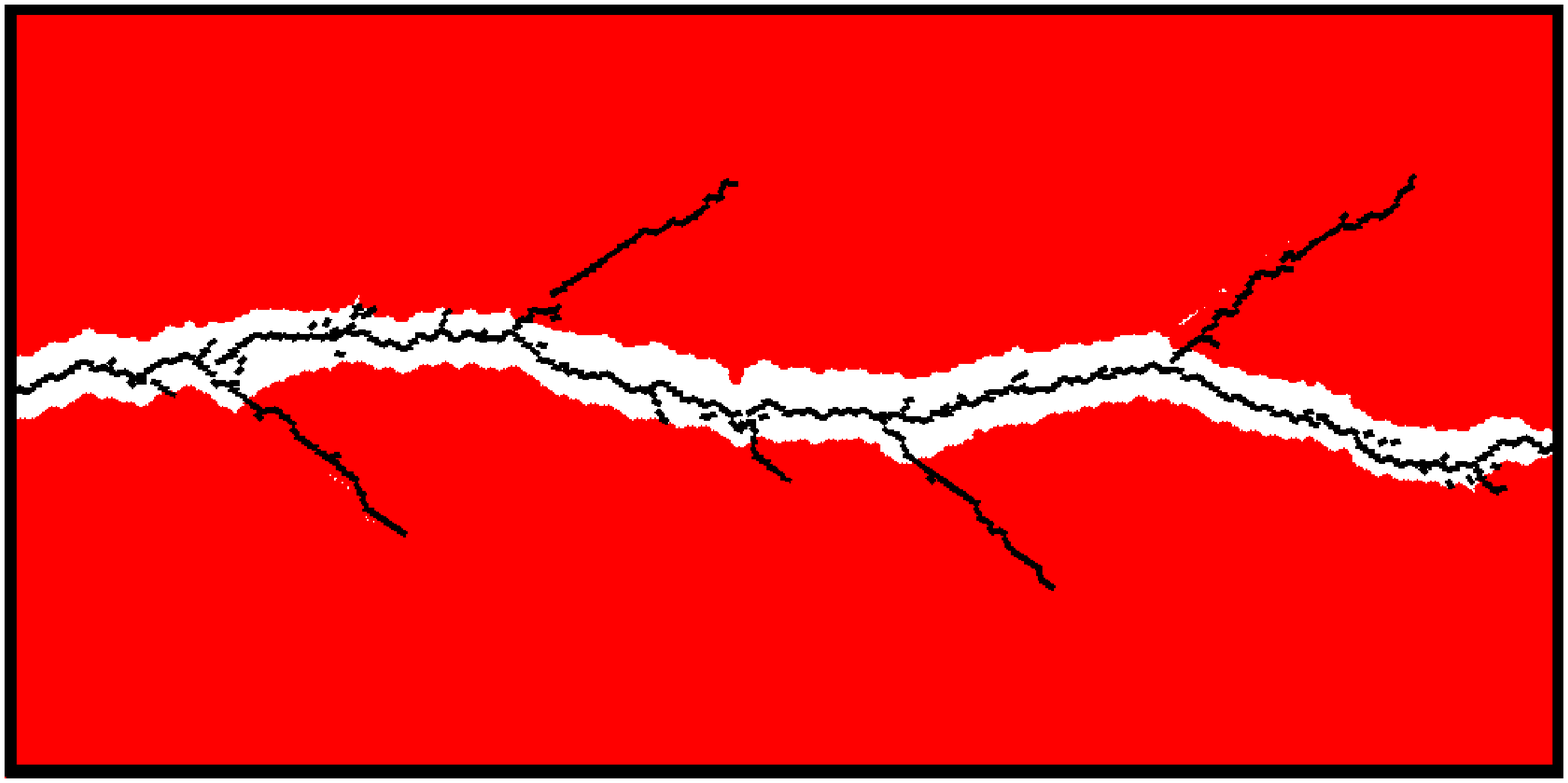}
}
\caption{(a) The pattern of the final cracked lattice using a perturbed honeycomb lattice model using $b=10\%$, $\eta=2$ and $\nicefrac{\Delta}{\Delta_G}=3.8$. (b) Same with
$\eta=2.5$ and $\nicefrac{\Delta}{\Delta_G}=4$.}
\label{beehive_crack2}
\end{figure}

Using a finite value of $b$ to perturb the lattice, we obtain nice snapshots of micro-branches, very much like those obtained using the CRN~\cite{shay3}. In Fig. \ref{beehive_crack1}(a) we
can see that when the driving displacement exceeds some value, a large micro-branch starts to appear. In Fig. \ref{beehive_crack2} we can see the final pattern
of broken bonds for two cases, one for $\eta=2$ (Fig. \ref{beehive_crack2}(a)) and one
for $\eta=2.5$ (Fig. \ref{beehive_crack2}(b)). The patterns looks very much like the fracture pattern seen using the CRN. Moreover, the
micro-branches also look similar to the experimental images of micro-branches in
PMMA~\cite{fineberg_sharon0,fineberg_sharon1,fineberg_sharon2,fineberg_sharon3,fineberg_sharon4,fineberg_sharon5}. This is an important result. As far as we know, this
is the first time that such a micro-branch pattern has appeared in a {\em lattice} material (previously shown by us only using an amorphous material model~\cite{shay3}).
We note that when the main crack continues and
the micro-branch arrests, one piece of the lattice overlaps with another piece (see Fig. \ref{beehive_crack1}(b)). This is a non-physical effect and is caused from the
fact that cracking in this model is irreversible. This effect is not large using the honeycomb lattice, but will be much more pronounced for the hexagonal lattice.

\begin{figure}
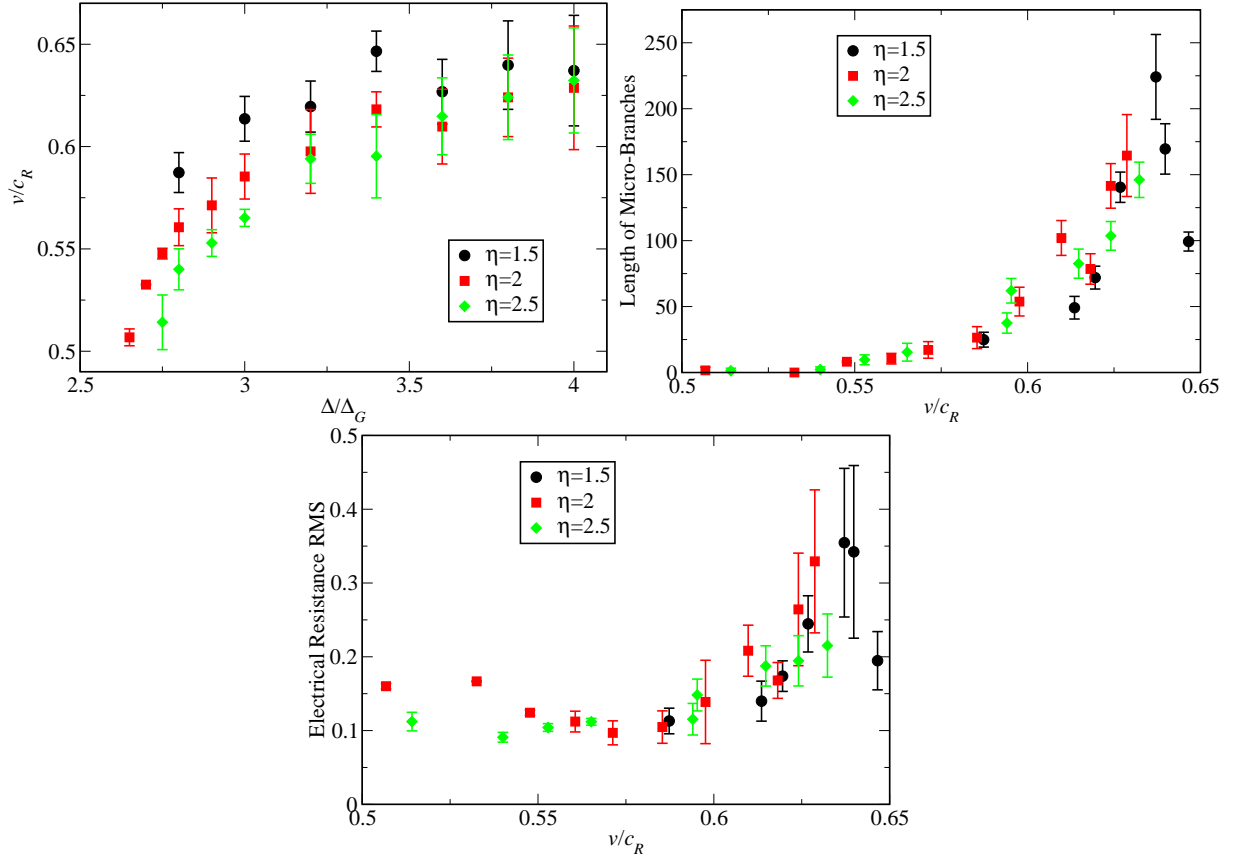

\centering{
\includegraphics*[width=8cm]{delta_v_10_ahuz.eps}
\includegraphics*[width=8cm]{v_micro_10_ahuz.eps}
\includegraphics*[width=8cm]{v_rms_10_ahuz.eps}
}
\caption{(a) The resulting $v(\Delta)$ curve for a perturbed honeycomb lattice model using $b=10\%$ with different $\eta$. The error bars were calculated using several simulations, each with
small change in $dt$, due to the chaotic nature of the problem (b) Size of total number of micro-branches as a function of the crack velocity $v$. For small velocities the total size
of micro-branches tends to zero (a steady state crack). (c) The amplitude of the oscillations in the electrical resistance as a function of the crack velocity. Even in steady-state cracks,
the amplitude of the oscillations tends to a certain finite (non-zero) value.
}
\label{beehive_results}
\end{figure}

\begin{figure}
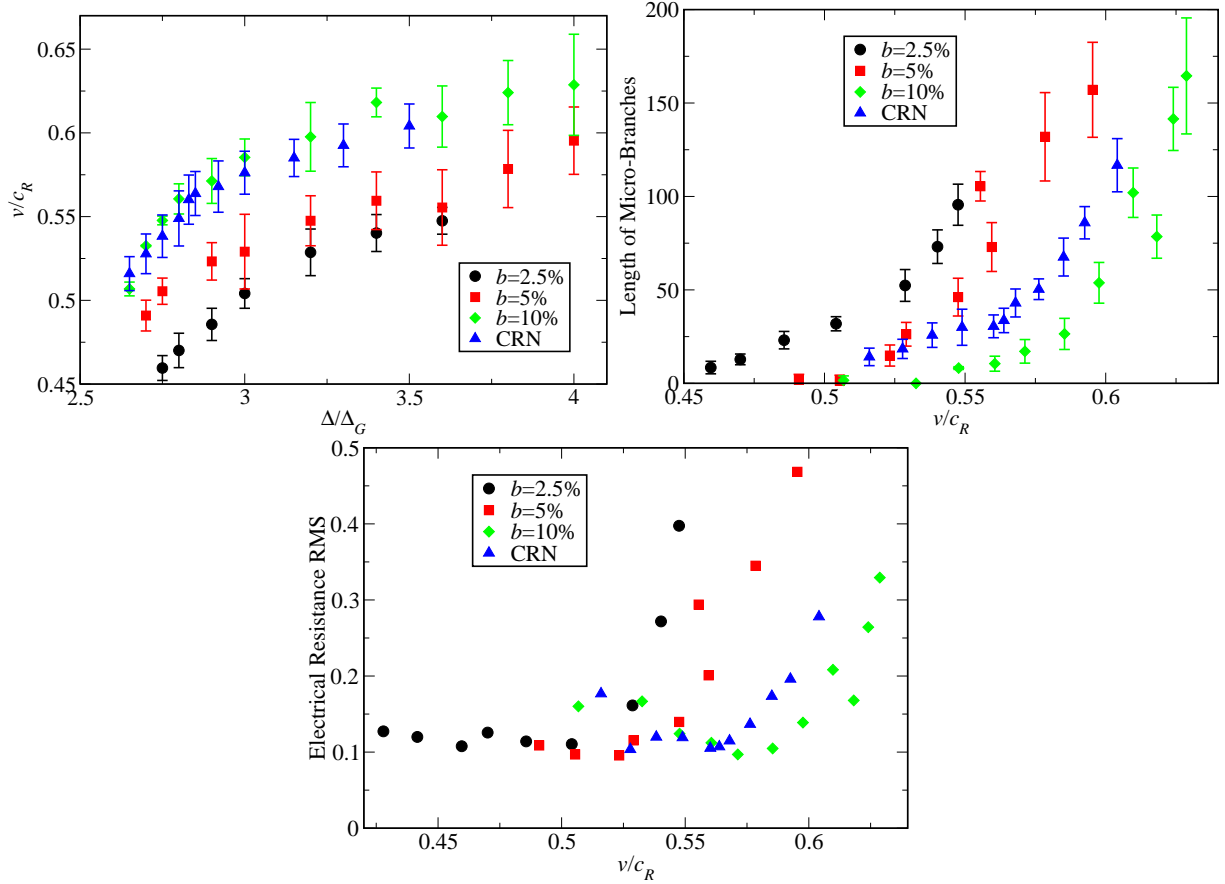

\centering{
\includegraphics*[width=8cm]{delta_v.eps}
\includegraphics*[width=8cm]{v_micro.eps}
\includegraphics*[width=8cm]{v_rms.eps}
}
\caption{(a) The resulting $v(\Delta)$ curve for a perturbed honeycomb lattice model with different values of $b$, along with the CRN results (taken from~\cite{shay3})
using $\eta=2$. The CRN curve is closer to the $b=10\%$ curve (b) Size of total number of micro-branches as a function of the crack velocity $v$. The somewhat questionable transition
between steady-state cracks and micro-branches behavior using the CRN, looks much more clear using perturbed lattices models. (c) The amplitude of the oscillations in the electrical
resistance as a function of the crack velocity. Again, the CRN curve is closer to the $b=10\%$ curve.
}
\label{beehive_results2}
\end{figure}

Beyond obtaining  the qualitative features of the micro-branches, the perturbed lattice model reproduces the main semi-quantitative results of mode-I fracture beyond the onset of instability.
Because of the chaotic nature of the problem (small changes in the simulation parameters yield different crack patterns, but with similar quantitative properties, such as the crack velocity),
we changed the time-step by a little bit ($\pm15\%$) for each driving displacement $\Delta$, to have sufficiently good statistics on the resulting parameters (about 20 runs for each point, therefore
each point in Fig. \ref{beehive_results}(a-c) represents a set of $\approx20$ runs). 
In Fig. \ref{beehive_results} we can see the $v(\Delta)$ curve (a), the total length of the micro-branches (b) and the amplitude of the oscillations of the electrical resistance (c).
The $v(\Delta)$ curve looks very much like the typical $v(\Delta)$ curve for mode-I fracture (for example, see~\cite{shay1,shay2,shay3}). The error bars represent the statistical error
of the crack's velocity using several simulations for each $\Delta$. The graph of the total length of the micro-branches as a function of the crack's velocity is one of the most
important results of this work. We can easily see that the total length of micro-branches goes to zero at small velocities, with a clear growth (near $v\approx0.57c_R$) with increasing velocity
(or $\Delta$).
This is in direct accord with the to experiment results (but instead of a sharp transition between steady-state cracks area and micro-branches area, we get a smooth transition, due to the noisy character of
discrete atomistic simulation at smaller scales). Actually, the experimental results  refer as the length of an average single micro-branch,
but because the lack of statistics (we have only a few micro-branches in each
single simulation, so it is hard to define the length of an average single micro-branch), we use the total sum of broken bonds instead (besides of course the main crack). This should be a
sufficiently close substitute (we see that we {\em do not} have just more ``short" micro-branches at large drivings; the size of each micro-branch indeed grows).
Comparing to the corresponding result using the CRN for amorphous material~\cite{shay3} (see also the triangles in
Fig. \ref{beehive_results2}(b)), where this transition was less pronounced, here it is much more clear. We note that by defining the total size of the micro-branches we subtracted all
the ``micro-branches" of size  1 or 2 broken bonds, which we neglect and treat as numerical noise.

From Fig. \ref{beehive_results}(b) we can see that using $\eta=1.5$ yields a too noisy system and we get either arrested cracks or significant micro-branches. There is no intermediate
zone of steady-state cracks. Only using $\eta=2$ do we yield clear steady-state cracks. Increasing $\eta$ further more to $\eta=2.5$ does not change the results appreciably.
In addition, the amplitude of the oscillations of the electrical resistance shows a nice agreement with the experimental result, as well as the CRN results~\cite{shay3} (For a wide
discussion in the different terminology between the crack's velocity oscillation and the electrical resistance oscillations and the appropriateness of the  ``electrical resistance" oscillations as a diagnostic,
see~\cite{shay3}). Beyond the ``critical velocity"
the amplitude of the oscillations increases rapidly, while for small velocity the amplitude of the oscillations is constant.

The sensitivity to the value of $b$, characterizing the width of the bond length distribution, was also investigated, exploring how much we can reduce $b$ and still get a physical behavior of micro-branches pattern, recalling that
$b=0$ (pure lattice) {\em does not}
create a physically realistic pattern. For $b\leqslant 1\%$, we reproduce the non-physical perfect lattice behavior, i.e. the perfect lattice behavior is not a singular case, in the sense that infinitesimal
change in the lattice produce micro-branches. You need a significant perturbation to yield physical behavior. In fact, even for $b=2.5\%$, the micro-branches seems to be in an almost
straight lines, along the preferred lattice directions, and thus, less physical. In Fig. \ref{beehive_results2} we can see the quantitative results using different values of $b$, along
with the CRN results (taken from~\cite{shay3}).

From the $v(\Delta)$ curve (Fig. \ref{beehive_results2}(a)) and the amplitude of the oscillations (Fig. \ref{beehive_results2}(c)) we can see that the CRN is closer to the $b=10\%$ results.
In Fig. \ref{beehive_results2}(b) we see that low $b$ results look like the CRN result. This fact
encourages us to conclude that the transition of the CRN results between low velocities and high is real, since increasing $b$ we get the same effect, but with a much sharper transition. 

\section{Hexagonal Lattice}

\begin{figure}
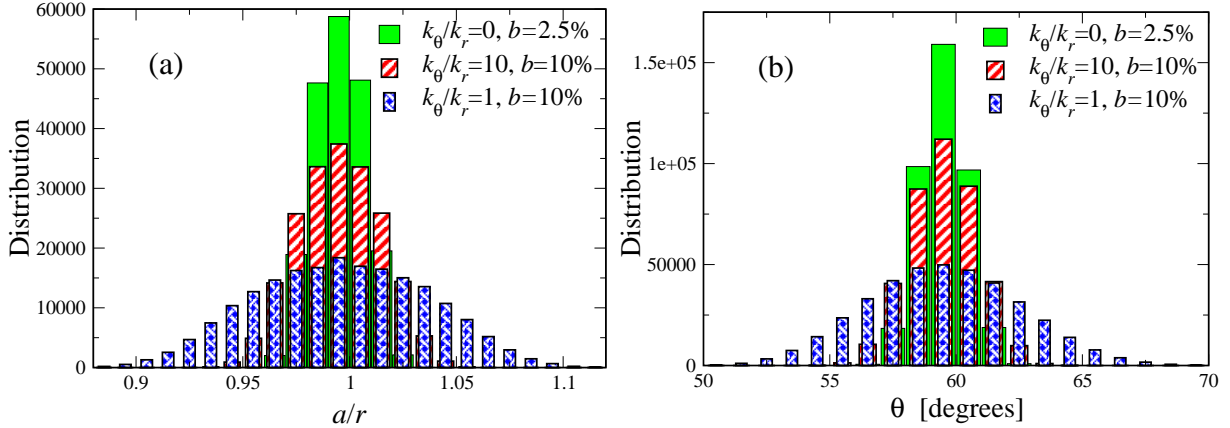

\centering{
\includegraphics*[width=8cm]{r_dist_hex.eps}
\includegraphics*[width=8cm]{teta_dist_hex.eps}
}
\caption{The radial (a) and the angular (b) distributions for a perturbed hexagonal lattice mesh with and without 3-body force law using different values of $b$.
$k_{\theta}/k_r=1$ and $b=10\%$ yields the almost the same distributions as $k_{\theta}/k_r=0$ and $b=10\%$ does.
}
\label{hex_dist}
\end{figure}

The classic models concerning molecular dynamics fracture simulations in perfect lattices used a hexagonal
lattice~\cite{marderliu,mardergross,pechenik,fineberg_mar,shay1,shay2,mdsim2,nature_old,nature}. As mentioned in the Introduction, those models were able to show nice qualitative results in the
mode-III fracture simulations~\cite{marderliu,kess_lev3}, but failed to create the physical
pattern of micro-branches in mode-I fracture simulations, investigated in the experiments~\cite{fineberg_mar,shay1}. The new findings from the previous section, that a small perturbation in the
potential between each two atoms {\em can} create a physical pattern of mode-I fracture, suggest it is worthwhile to try it also in the classic hexagonal lattice.
The resulting radial and angular distributions for several parameters are shown in Fig. \ref{hex_dist}.

\begin{figure}
\centering{
\includegraphics*[width=7cm]{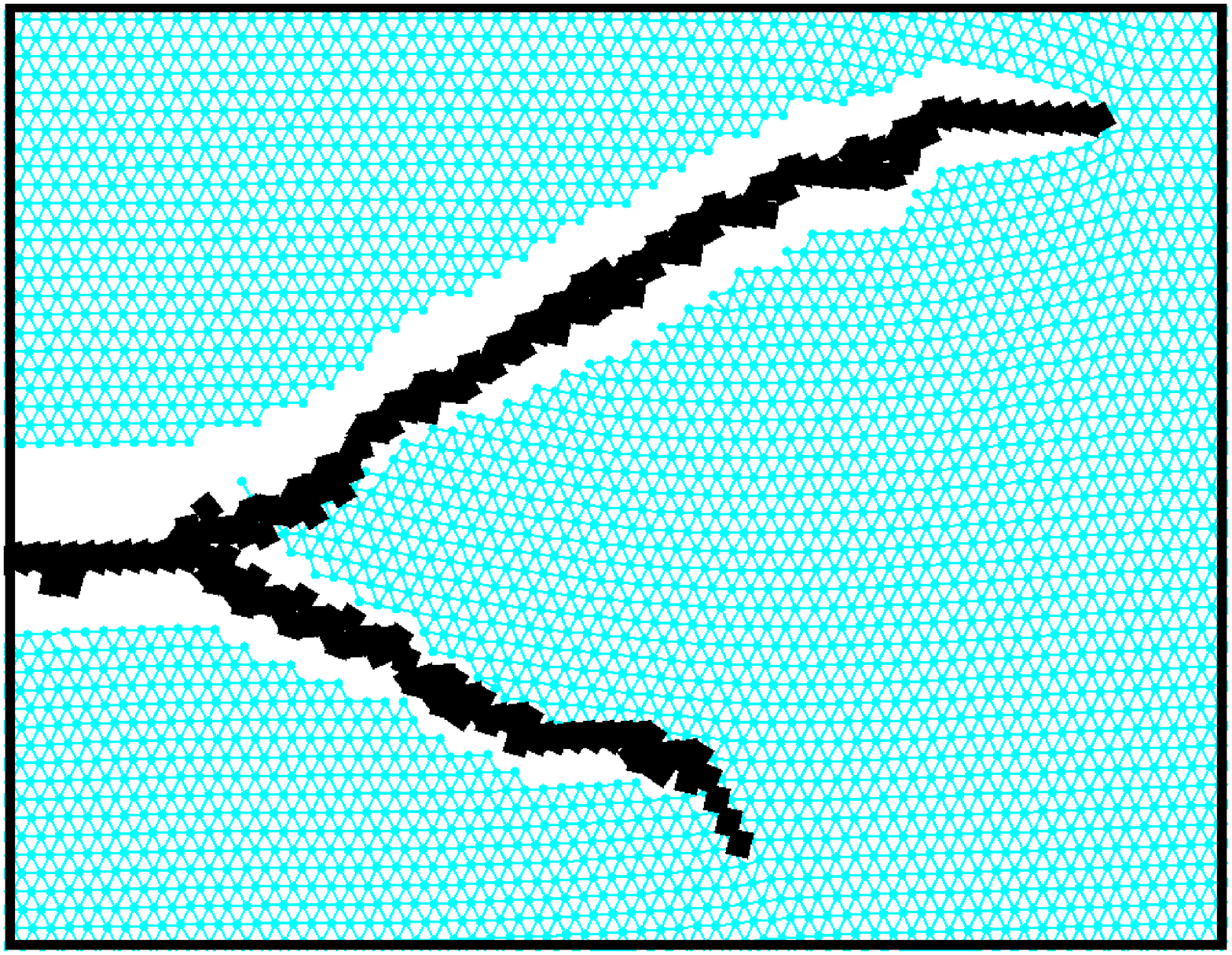}
\includegraphics*[width=6.8cm]{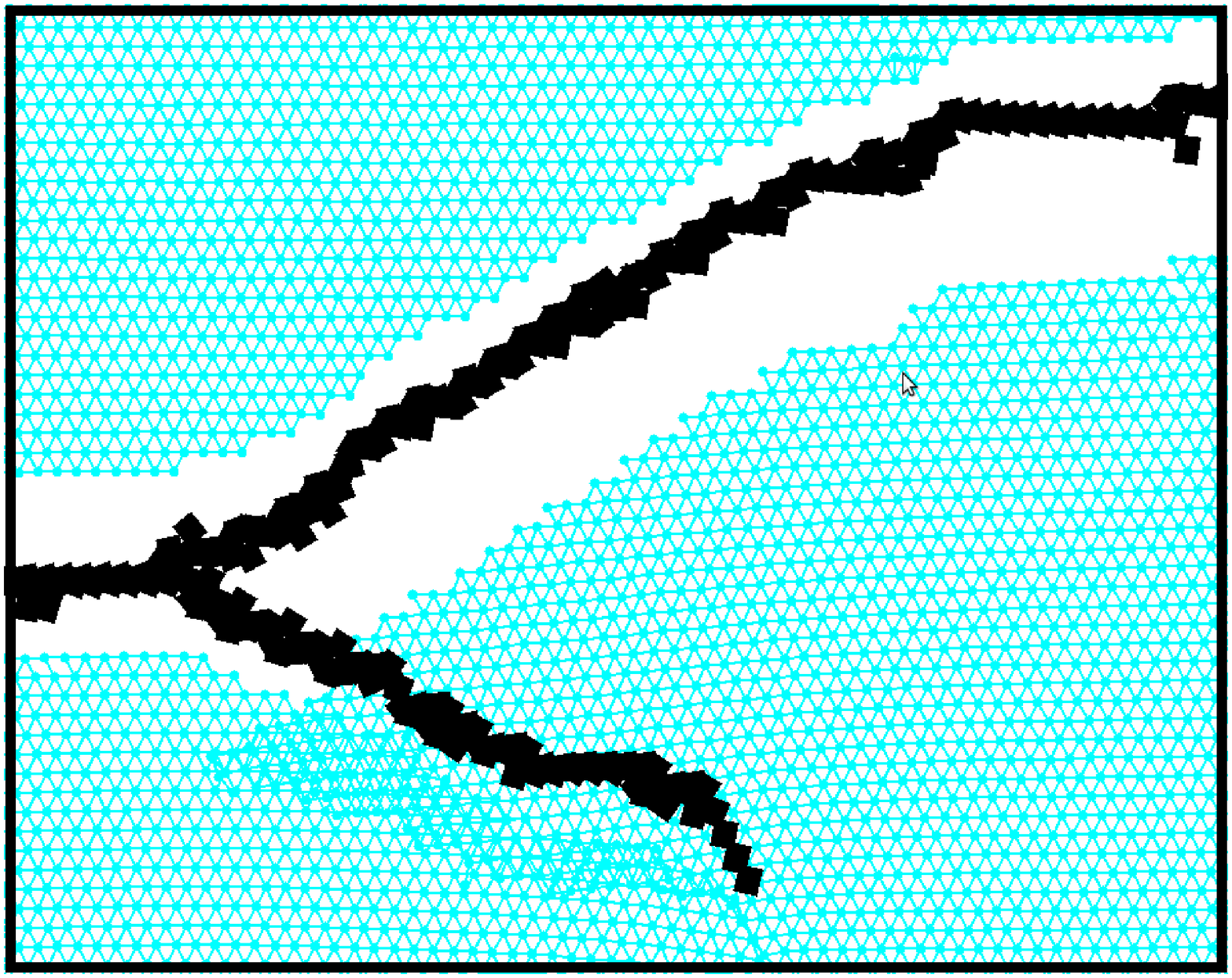}
}
\caption{(a) A propagating crack snapshot using a perturbed hexagonal-lattice model using $b=2.5\%$, $\eta=0.25$ and $\nicefrac{\Delta}{\Delta_G}=2.2$. (b) A snapshot of an extensive
overlapping problem. After cracking, the two pieces of the cracked lattice overlap, because fracture is irreversible process in this model. 
}
\label{hex_crack1}
\end{figure}

\begin{figure}
\centering{
\includegraphics*[width=7cm]{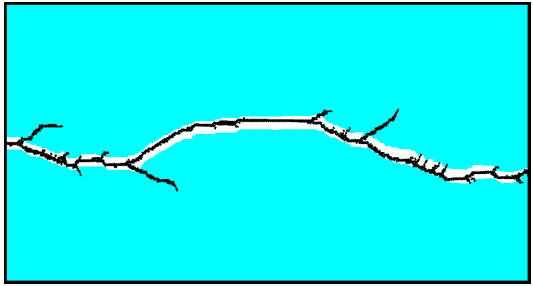}
}
\caption{The pattern of the final cracked lattice using a perturbed hexagonal-lattice model using $b=2.5\%$, $\eta=0.25$ and $\nicefrac{\Delta}{\Delta_G}=2.2$.}
\label{hex_crack2}
\end{figure}

Using the same magnitude of perturbation as in the honeycomb lattice, $b=10\%$ (with no 3-body force law, i.e. $k_{\theta}=0$) yields an extreme noisy simulations, and eventually, in most cases,
the crack arrests. Using a smaller value of $b$ ($b=2.5\%$), yields in general micro-branching patterns (for large driving displacement)
(Figs. \ref{hex_crack1} and \ref{hex_crack2}), but it still looks less physical than the honeycomb lattice results. At low drivings, there is a single steady-state crack propagating in the midline
of the sample yielding no broken bonds besides the main crack.
Nevertheless, the benefit of using this model (in the absence of a 3-body force law) is that we can compare it to Slepyan's lattice steady-state models (like in~\cite{shay1}, using large
$\alpha$, which corresponds to a piecewise-linear model), when the origin of instability
between steady-state cracks and micro-branches behavior is known exactly (although the models use $b=0$, the small perturbation does not change the results significantly). Using $b$ smaller than
$2.5\%$ reproduces the non-physical behavior of perfect lattices. In addition, the hexagonal mesh allows us to work with small values of $\eta$ which is more
relevant experimentally~\cite{livne,bouchbinder_exp}, than the large $\eta$ honeycomb lattice.

\begin{figure}
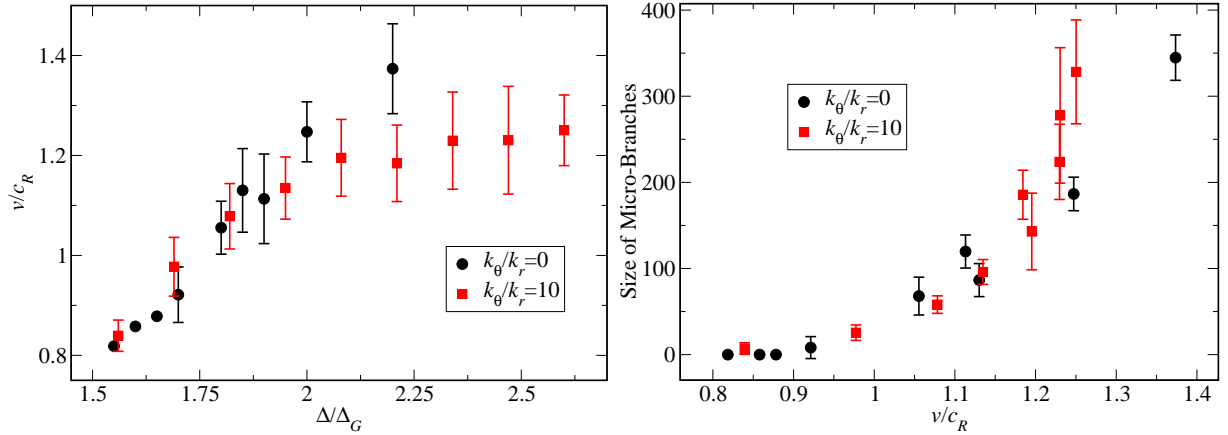

\centering{
\includegraphics*[width=8cm]{delta_v_hex.eps}
\includegraphics*[width=8cm]{v_micro_hex.eps}
}
\caption{(a) The resulting $v(\Delta)$ curve for a perturbed hexagonal-lattice model using different values of $k_\theta$ with $\eta=0.25$, ($b=2.5\%$ for $k_\theta/k_r=1$ and $b=10\%$ for $k_\theta/k_r=10$.$b=10\%$. The error bars were calculated using several simulations, each with
small change in $dt$, due to the random nature of the problem (b) Size of total number of micro-branches as a function of the crack's velocity $v$. In small velocities the total size
of micro branches tend to zero (a steady state crack). The including 3-body results are normalized to the non 3-body results. 
}
\label{hex_results}
\end{figure}

In Fig. \ref{hex_crack1}(a) we can see the a birth of a micro-branch that bifurcates from the main crack, that eventually arrests. The pattern looks very much alike the honeycomb lattice.
In Fig. \ref{hex_crack2} we can see the final pattern of broken bonds for the case of $\eta=0.25$. The quantitative results regarding the $v(\Delta)$ curve and the total size of
micro-branches as a function of $v$ are presented in Fig. \ref{hex_results}.

We can see in Fig. \ref{hex_results}(a) that the velocities below $v=c_R$ reproduce the Slepyan's lattice model results from~\cite{shay1}, yielding perfect steady-state cracks, with no
micro-branches at all (Fig. \ref{hex_results}(b)). Increasing the driving displacement $\Delta$ further, yields a non steady-state behavior, as the steady-state lattice model solution becomes
unstable, yielding a micro-branching behavior (again, only above a threshold $b$). We note again that in many cases the fracture pattern do not looks as physical as in
Fig. \ref{hex_crack2}, yielding some non-physical results. In addition, the problem of overlapping zones is more extensive in the hexagonal lattice (see Fig. \ref{hex_crack1}(b)), yielding
large areas of overlapping zones and thus, yielding an unphysical behavior. The main benefit so far of using this model is that Fig. \ref{hex_results}(b) yields similar qualitative results to
Fig. \ref{beehive_results}(b) or Fig. \ref{beehive_results2}(b), emphasizing that the transition at low velocities to a steady-state behavior is real, since in a hexagonal case we
get absolutely zero micro-branches at low-velocities, and  agreement with the Slepyan's lattice models.

Adding a 3-body force law in a hexagonal lattice, one must use large values of $k_{\theta}$ since the relatively large number of nearest neighbors doesn't allow each 3 atoms to generate
an angle significantly larger than $\nicefrac{\pi}{3}$ using $k_{\theta}\approx k_r$ (as was the case in the honeycomb lattice), and thus, the 3-body energy is negligible. When we increase $k_{\theta}$
using $b=2.5\%$, we yield a perfect lattice behavior because $k_{\theta}$ is too large. Increasing $b$ further to $b=10\%$ yields a good balance which the 3-body energy is not negligible and
yields similar radial and angular distributions as a hexagonal lattice without the 3-body force law using $b=2.5\%$ (see Fig. \ref{hex_dist}).

We normalize the $v(\Delta)$ curve results and the total size of micro-branches as a function of $v$ using this model to the values of the non-3-body force law case (the normalization constants
$\Delta_G$ and $c_R$ are of course larger using large significant $k_{\theta}$) and present them in Fig. \ref{hex_results}. We can see that these two models share the same qualitative behavior
(although the including 3-body force law results yields a small number of micro-branches at low velocities, very much like the honeycomb case). In addition, this model still suffers from a severe
problem of overlapping of pieces of the mesh after branching (in contrast to honeycomb lattice, which in this model, this problem is minor).

\begin{figure}
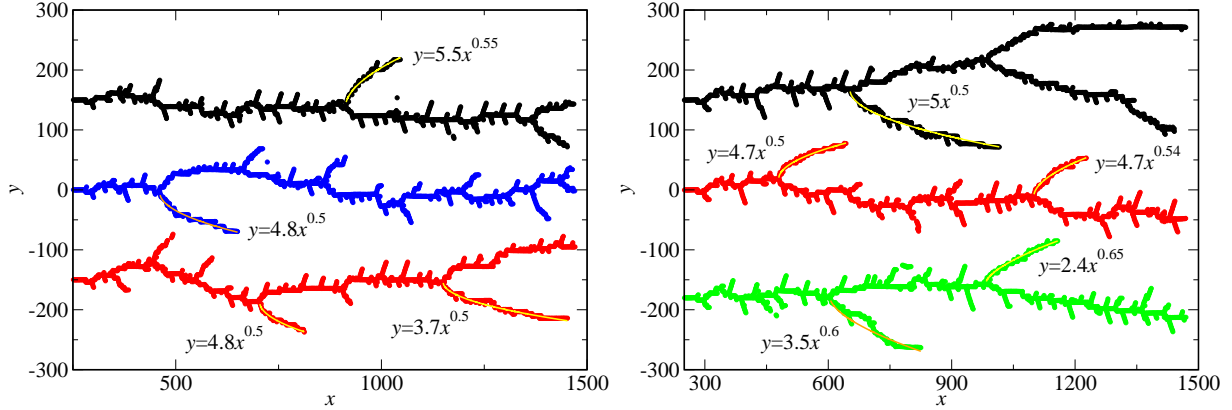

\centering{
\includegraphics*[width=8cm]{micro_power.eps}
\includegraphics*[width=8cm]{more_power.eps}
}
\caption{The crack pattern for a perturbed hexagonal-lattice model using $k_{\theta}/k_r=10$, $b=10\%$ and $\eta=0.25$. (a) $\nicefrac{\Delta}{\Delta_G}=1.8$ with different values of $dt$. (b) $\nicefrac{\Delta}{\Delta_G}=1.9$ with different values of $dt$ for the first two upper patterns, and $\nicefrac{\Delta}{\Delta_G}=2$ for the bottom pattern.
}
\label{micro_power}
\end{figure}

However, two surprising results appear using this model (a perturbed hexagonal lattice {\em including} 3-body force law). First, the main crack stays more confined to the middle of the sample,
even for large driving, with large micro-branches,
in contrast to all other models, including the honeycomb lattice model. Second, and most important, the larger micro-branches have a non-linear, power law shape (!) very much alike the experimental
data. In Fig. \ref{micro_power} we can see several of crack-patterns for different driving displacements (with different $dt$, with shifting in the $y$-axis, of course).
We add a basic power law fit (with different $x_0$ and $y_0$ for each large
micro-branch, of course). We got a nice power law behavior (but with power of $0.5-0.65$ instead of $\approx0.7$ of~\cite{fineberg_sharon0,fineberg_sharon2}). This is of course a very
preliminary result, and must be tested in larger scales. 
We note that the $\approx0.7$ power law behavior is not universal for all the experiments. At least in the experiments of~\cite{fineberg_sharon5}
(Fig. 3),~\cite{fineberg_sharon6} (Fig. 4(a)) and~\cite{nature_ko} (Fig. 1(b)), straight
micro-branches appear, so the physical behavior of Figs. \ref{beehive_crack2} and \ref{hex_crack2} is important too. On the other hand, power law behavior (different than 1) is seen
in the atomistic model only in a perturbed hexagonal lattice {\em including} 3-body force law. These observations support the conclusion that the macroscopic behavior of fracture depends strongly
on the inter-atomic microscopic potential. 

\section{Discussion}

We have shown that micro-branching  can be reproduced in lattice materials, using a small perturbation parameter $b$ which perturbs the inter-atomic potential between each
two atoms. In addition to the qualitative patterns of micro-branches, semi quantitative results are shown, particular the total size of micro-branches (which corresponds to the average
size of a micro-branch in the experiments) as a function of the velocity. A clear transition between steady-state behavior  and  the post-instability region is seen, characterized by
an increased number of broken bond off the midline of the sample. This result is in line with the results of the CRN model. In particular we obtain in the
hexagonal case (without 3-body force law) no micro-branching at all, as in the Slepyan lattice models. The increased amplitude of the RMS of the
electrical resistance is shown as well, in agreement to the experiments.

In addition, preliminary signs of  power law behavior for the shape of the side branches can be seen using a hexagonal lattice including a 3-body force law, in accord with experiment, and which has not been observed in previous atomistic models.
Much more extensive work is underway to extend this result to larger scales, as well as exploring the opening angle of the crack. 

\begin{acknowledgments}
The authors wish to thank David Srolovitz for useful discussions and remarks.
\end{acknowledgments}

\end{document}